\documentclass{ws-ijmpc}

\usepackage{graphicx}

\begin{document}

\markboth{Veit Schw\"ammle}
{Phase transition in a sexual age--structured model of learning foreign languages}

\catchline{}{}{}{}{}

\title{Phase transition in a sexual age--structured model of learning foreign languages}

\author{V.~Schw\"ammle}
\address{Instituto de F\'isica, Universidade Federal Fluminense, Av. Litor\^anea, 
s/n - Boa Viagem, 24210-340, Niter\'oi, RJ, Brasil.\\
Institute for Computer Physics, University of Stuttgart, Pfaffenwaldring 27, 
D-70569 Stuttgart, Germany.\\
veit@if.uff.br}

\maketitle

\begin{history}
\received{}
\revised{}
\end{history}



%

\begin{abstract}
The understanding of language competition helps us to predict extinction and survival of 
languages spoken by minorities. 
A simple agent--based model of a sexual population, based on the Penna
model, is built in order to find out under which circumstances one language dominates other ones. 
This model considers that only young people learn foreign languages. The simulations show a 
first order phase transition where the ratio between the number of speakers of different languages
is the order parameter and the mutation rate is the control one. 
\end{abstract}

\ccode{PACS Nos.:89.75.-k,89.65.-s}
  
\section{Introduction}
\label{sec:intro}


Recently, increasing attention has been paid on the understanding of linguistic systems 
by computational 
and analytical methods. Physicists, mathematicians, computer scientists and biologists 
apply their tools on the investigation of language capability, language change and language 
competition\cite{Wang2005}. 
Especially the similarity to biological systems opens the field of linguistics 
to methods of the far better understood evolutionary systems. For an overview see 
refs.\cite{Wang2005,Science2004,Stauffer2005}. 

It is believed that a concept of ``Universal Grammar'' is fixed by some way
in our genetic code enabling humans
to learn languages fast during our childhood\cite{Chomsky80}. 
Much attention has been paid on the evolution of this trait and the competition between 
different grammars\cite{Nowak2001,Komarova2004}. 
In these models the different grammars compete by giving higher fitness to 
better competitors. Because of that it is possible to apply directly models already 
successfully used in evolutionary biology. 

The competition between languages has been investigated in the last years analyzing the 
stability of an system of two or more 
languages\cite{Stauffer2005,Abrams2003,Patriarca2004,Kosmidis2005,Mira2005,Schwaemmle2005b} as 
well as their size distribution\cite{Schulze2005b,deOliveira2005}.
In our approach we use a computational model \emph{without} giving different fitness to 
agents with different language traits and without neglecting bilinguals. Models of that kind
can be found in refs.\cite{Stauffer2005,Mira2005,Schwaemmle2005b,Schulze2005b}.

This makes our model different from biological systems. The knowledge of a different
language does not imply a higher death probability (we are aware that, 
unfortunately, there are quite many exceptions of that rule).
In order to present an age--structured model we build our one on the Penna 
model\cite{Penna95,Stauffer2006}. By the treatment of the genes like bits the model yields 
a rather realistic age--structure of a sexually reproducing population, and  
is additionally suitable for large population sizes due to its low computational cost. 
The usage of an age--structured model enables us to merge an age--dependent learning procedure 
with a model of language competition in order get more insight into their 
relation. We will focus on the question whether a phase transition exists, 
as found in the models of refs.\cite{Stauffer2005,Komarova2004}, between the state 
of dominance of one language and an uniform distribution (also called fragmentation).

The paper is organized as follows: the next section explains the model, the following
section  presents the results of the model
with respect to its parameter space. The last section discusses the results and proposes
further extensions of the model. 


\section{The model}
\label{sec:model}

The model in this paper combines population genetics with age--dependent language learning. 
The presentation of the genetical part, with the purpose
to yield a stable population having its characteristic age--structure, will be provided only 
shortly to the reader. For a more detailed description of the Penna model and its 
applications we refer to refs.\cite{Penna95,Stauffer2006}. 

\paragraph{The sexual Penna model} is a individual--based model where every agent is 
represented by its age and two strings (diploid) of 32 bits. The bit--strings are read 
in parallel. Every iteration after birth a new position on each of the  
bit--strings becomes visible. 
A bit set to one corresponds to an harmful allele. At the beginning of a simulation five 
positions are randomly chosen to be dominant ones. If the position is dominant one bit of 
the two bit--strings suffices to make the position represent a deleterious mutation. On the other 
positions both bits need to be set to one for deleterious effects. 
At the age at which an agent has its third 
deleterious mutation it dies. In order to avoid population explosion an additional factor 
limits its growth: the Verhulst factor $1-N/K$ gives the probability of an agent to survive an 
iteration. Here $N$ is the current population size and $K$ the so--called carrying capacity 
defining the maximum size of a population. After reaching the reproduction age $R=10$
a female agent every iteration chooses randomly a male with age equal or older than $R$ 
in order to produce one offspring. 
For each parent its two bit--strings are randomly crossed and recombined. Both parents contribute 
one string to the offspring. The offspring suffers one deleterious mutation 
at a randomly chosen position on each of its both new strings. This high 
mutation rate lets the population reach a stationary distribution fast.  
The two bit--strings are initialized randomly. 

\begin{figure}[htb]
  \begin{center}
    \includegraphics[width=0.4\textwidth]{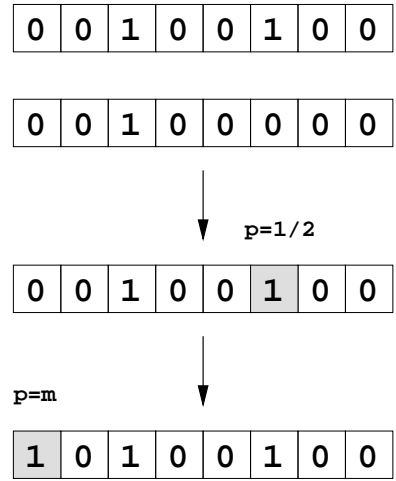}
\caption{The language string of the child is built by the two parent's language strings. 
If both parents speak a language, the child speaks it as well. If only one of the two 
parents speaks one, the child speaks it with 50\% probability. After composition of the 
child's bit--string mutations with probability $m$ generate a new language spoken by the 
child on a randomly chosen position.}
    \label{fig:illustrate}
\end{center}
\end{figure}

\paragraph{The language trait} of an agent, that means its current ability to speak certain 
languages, is modelled in a similar way as the genetic strings of biological ageing. 

The whole structure, a language should consist of, is neglected, and language is treated as 
an unit. Each agent contributes one additional bit--string 
of $L$ bits representing its capability to 
speak a maximum of $L$ languages. So an agent who has for instance its third bit set, speaks 
currently language $l=3$ (eventually among others). 
An agent can learn or forget a language only in youth during the first $c$ 
iterations after birth. We call $c$ the \emph{maximum learning age}. Older agents are not 
able to change their knowledge on languages. 
The interaction between a young agent and a randomly chosen teaching agent works 
as follows: At first a random position on the language bit--string is chosen. The agent 
learns the language of that position if it is spoken by the teacher. If the agent already 
speaks more than one language it forgets the language of the chosen position if the teacher 
does not know it. In the other cases no process of learning/forgetting is carried out. 
Let us illustrate the outcome of the interaction in computational terms: 
the language trait of an agent speaking zero or one 
language has an OR operation with the teacher's one at the chosen position, otherwise an AND 
operation. The young agents's language traits are actualized in this way by 
$f$ different teachers at different positions per iteration. $f=0.5$ means that the language
traits are actualized once per iteration with a probability of $50\%$.
At birth an offspring obtains its language trait as illustrated in 
Figure~\ref{fig:illustrate}. If both parents speak the same language the offspring will 
speak it as well. In the case that only one of the parents speaks a language the child 
learns it with a probability of $50\%$. Additionally, the child can know a new language at 
birth  with probability $m$ (i.e. a randomly chosen bit position is set to one). We call 
this probability $m$ of erroneous learning the \emph{mutation rate} in the following because of 
its analogy to biological systems. The bit--string is initialized by chance at the 
beginning of a simulation.

\section{Results}
\subsection{Two languages}
\label{sec:res}

\begin{figure}[htb]
  \begin{center}
    \includegraphics[angle=270,width=0.8\textwidth]{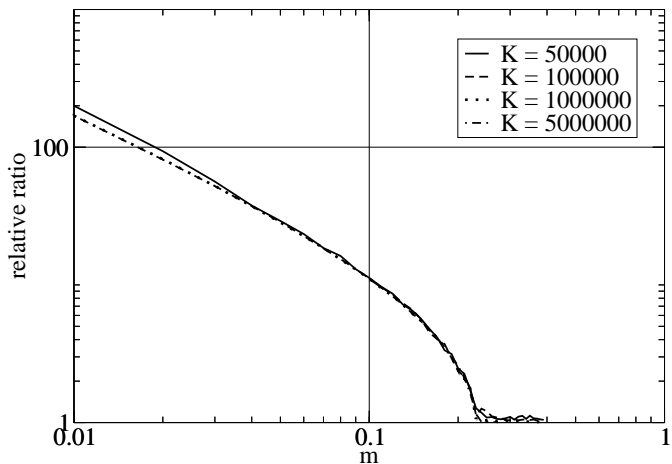}
\caption{Ratio of the two language sizes versus mutation rate. The size is the number of agents
speaking that language. The figure shows a clear phase transition. 
Different population sizes alter neither the shape nor the position of the critical point.}
    \label{fig:phasetr1}
\end{center}
\end{figure}

This section will concentrate at first on the results of simulations with a trait of
two languages ($L=2$). The simulations are carried out for at least $20,000$ iterations 
or more for large population sizes. All results present the final stationary state of a 
simulation.

\begin{figure}[htb]
  \begin{center}
    \includegraphics[angle=270,width=0.8\textwidth]{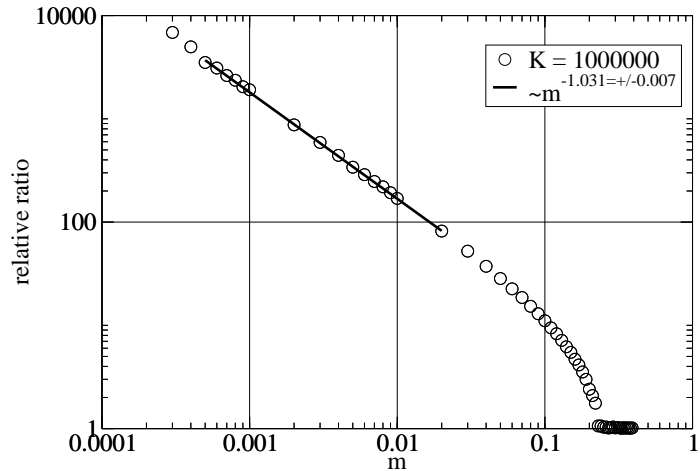}
\caption{The language ratio shows a power law with an exponent of about minus one for small 
mutation rates.}
    \label{fig:powerlaw}
\end{center}
\end{figure}

A phase transition can be observed for the ratio between the number of 
speakers of different languages. The ratio is averaged over the last $1,000$ time steps.
Figure~\ref{fig:phasetr1} shows the ratio versus the mutation rate, determining the control
parameter,
for different carrying capacities $K$. The simulations are carried out with a maximum 
learning age of $c=4$ and $f=0.5$. Each point represents the outcome of a separate simulation. 
The language ratio decreases fast at the critical point which is situated between $m=0.2$ and
$m=0.3$. The transition separates the phase/state of about equal sizes for the two 
languages for small values of $m$
from the phase/state where one language dominates clearly the other one. 
We can observe in Figure~\ref{fig:phasetr1} as well that the population size does not alter
the shape of the transition. Thus this transition can be found for arbitrary population numbers 
and therefore corresponds to a real physical phase transition.

The language ratio increases strongly for low mutation rates. A power law with an
exponent of minus one is exhibited in Figure~\ref{fig:powerlaw}. 
This exponent is the same for all simulations presented here. The parameters in 
the simulations are $c=4$, $f=0.5$ and $K=1000000$, i.e. 
the population consists of about $80,000$ agents.  

\begin{figure}[htb]
  \begin{center}
    \includegraphics[angle=270,width=0.8\textwidth]{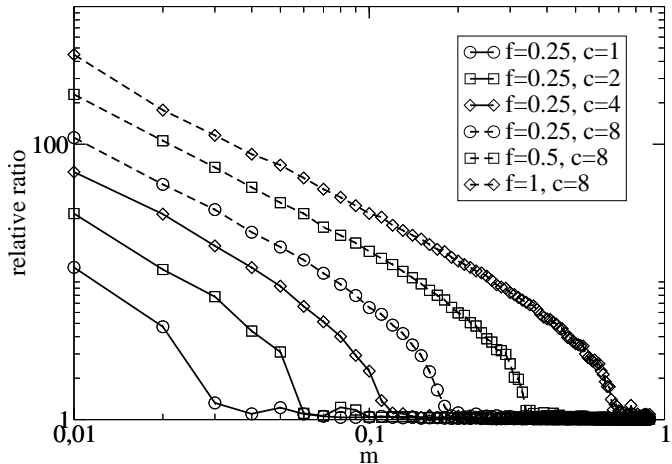}
\caption{The curve of the phase transition is shifted to large mutation values for increasing 
learning time during childhood. The shift increases linearly with an increasing maximum 
learning age $c$ as well as with increasing amount $f$ of interaction per iteration.}
    \label{fig:phase_interact}
\end{center}
\end{figure}

In Figure~\ref{fig:phase_interact} the shape of the phase transition is compared for different 
values of
$c$ and $f$, parameters defining the amount of interaction between the agents.
The carrying capacity $K=100,000$ fixes the population number at around $8,000$.
The shape of the transition remains the same but is shifted to higher values of the  mutation 
rate for increasing $c$ and $f$. The shift is a linear one with respect to each of both 
parameters. The shift decreases for values of $c>10$ as expected due to the 
smaller number of individuals surviving up to ages much older than the reproduction age $R=10$ 
(not shown). The phase transition disappears and the only final state of the model is the dominant
one for very large values of $f$.


\subsection{Many languages}
\begin{figure}[htb]
  \begin{center}
    \includegraphics[angle=270,width=0.7\textwidth]{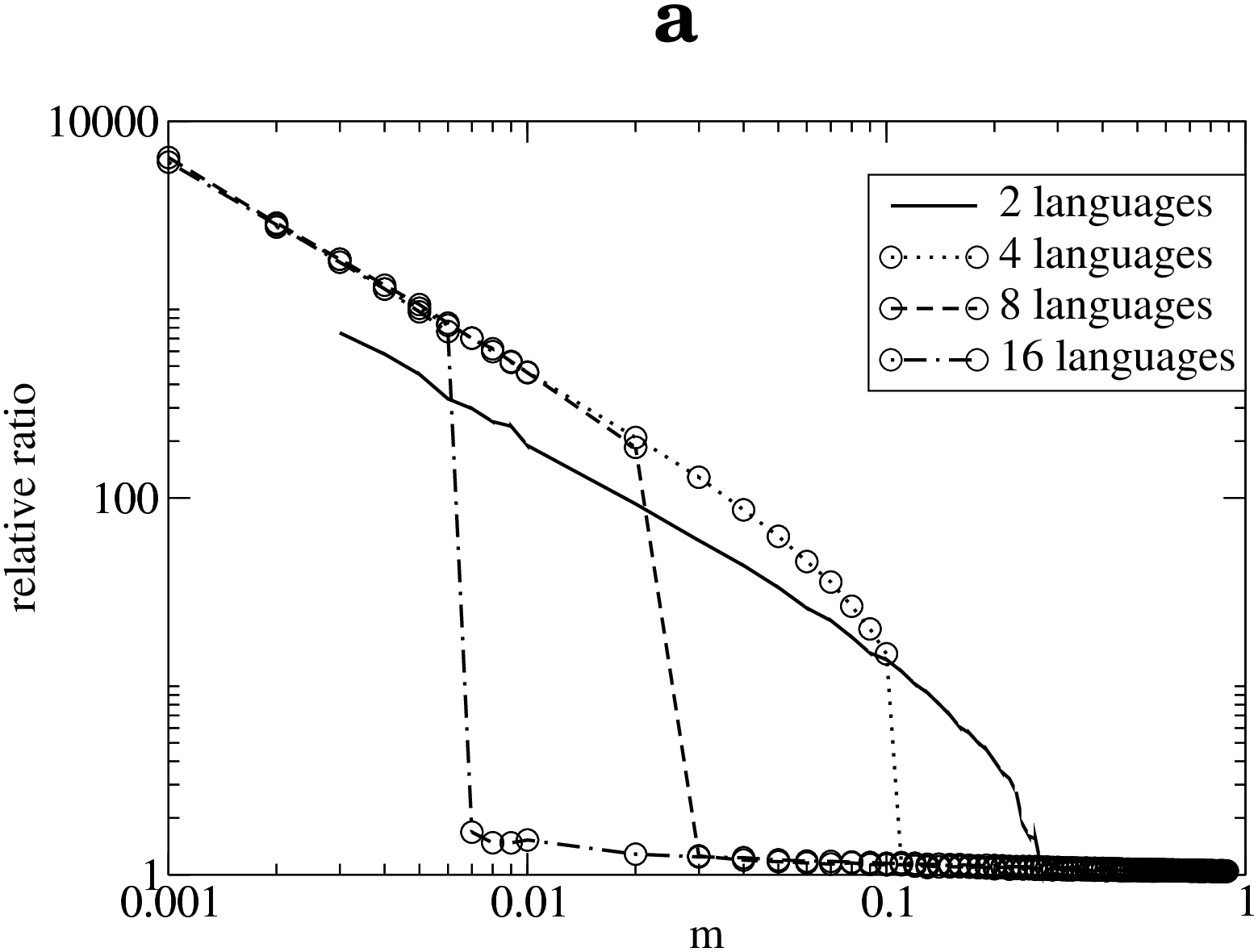}
    \includegraphics[angle=270,width=0.7\textwidth]{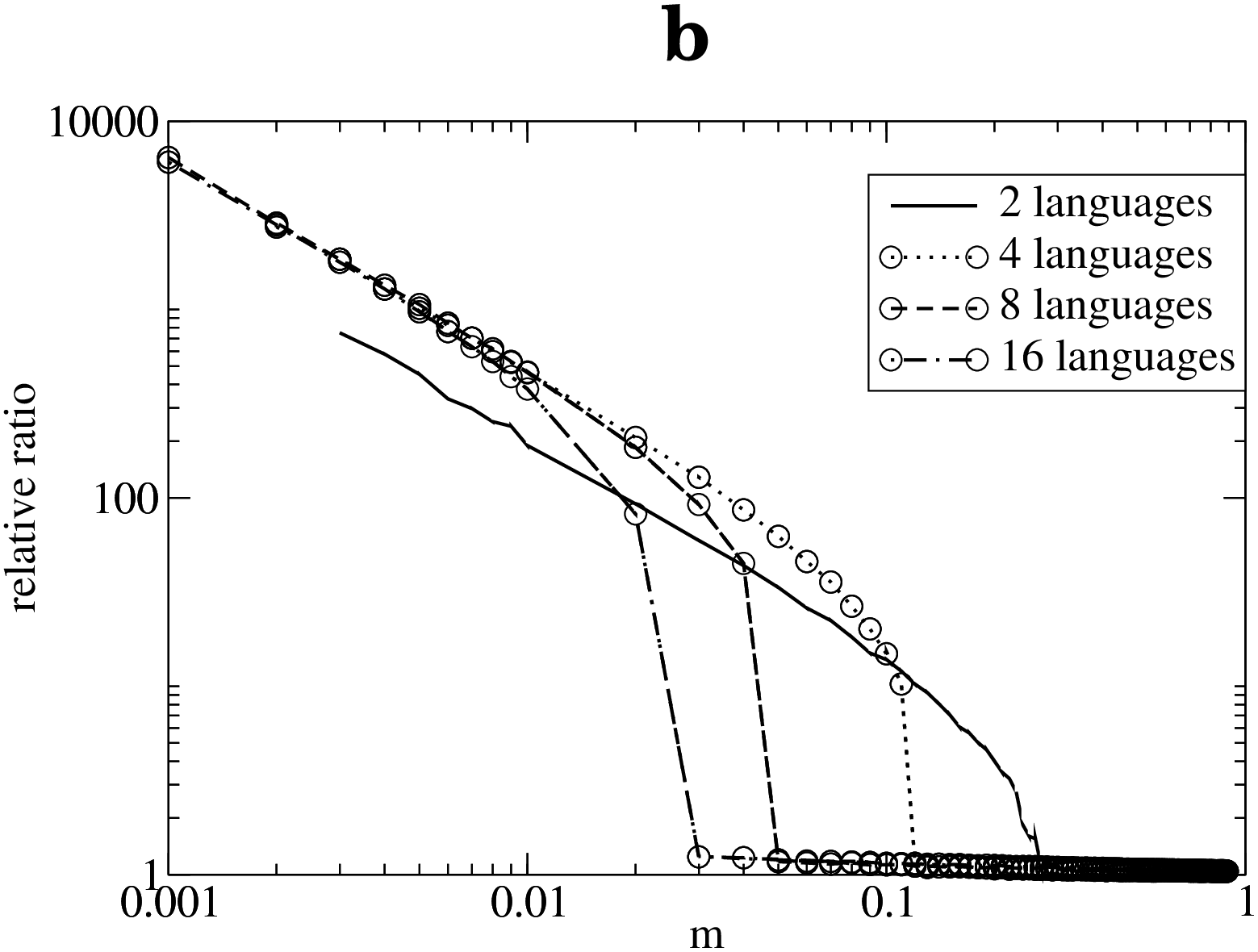}
\caption{Ratio of languages versus mutation rate for different numbers of $L$ languages. The 
critical point moves to smaller values of the mutation rate for increasing $L$. 
The phase transition is more abrupt. 
{\bf a}: For numbers of languages larger than two the curves collapse into one for values of $m$
in the left part where 
one language dominates the other ones. {\bf b}: The initialization with one language leads to 
a different result. The transition shows a hysteresis for the simulations with more
than two languages increasing with $L$.}
    \label{fig:numwords}
\end{center}
\end{figure}

In the following the results of simulations with a language trait consisting of more than 
two languages are compared to the previous ones. The language ratio $R$, the measure 
for a possible domination of one over the others, is now defined by 
\begin{equation}
R = \frac{N(l_{max}) \cdot (L-1)}{\sum \limits_{l=1, l\neq l_{max}}^{L} N(l)},
\end{equation}
where $l_{max}$ is the most spoken language, $L$ the number of languages, and $N(l)$
the number of agents speaking language $l$. The simulations are carried out with $c=5$ and 
$K=100,000$. We used $f=0.5,1,2,4$ for $L=2,4,8,16$, respectively in order to keep
the interaction per language constant. 
Figure~\ref{fig:numwords}a compares the shape of the transition for different numbers 
of languages. The position of the critical point is shifted to smaller
values of $m$ for more languages. This means that the more languages 
compete against each other, the lower is the possibility to have a scenario where one language 
dominates the other ones. 
Figure~\ref{fig:numwords}b compares the same simulations carried out with the initial state
where all agents speak language $l=1$. The critical point is shifted to 
larger values of the mutation rate. This dependence of the final state on the initialization, 
leading to a hysteresis, shows that the phase transition is of first order. Here the jump from
one phase into the other is located at the intersection of the transition curve 
and the curve for the ratio of two languages.

\begin{figure}[htb]
  \begin{center}
    \includegraphics[angle=270,width=0.8\textwidth]{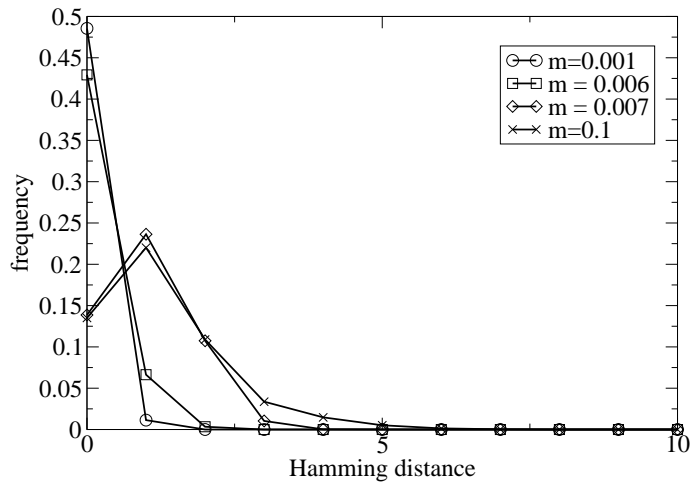}
\caption{Histogram of the Hamming distances between the agents in a simulation of 16 languages 
for different values of the mutation rate.}
    \label{fig:hamming}
\end{center}
\end{figure}

Finally, we present a histogram of the Hamming distances\cite{Tesileanu2006} 
between all agents for simulations with 16 languages for different values of the mutation rate 
(Figure~\ref{fig:hamming}). The parameters are the same as before.
The Hamming distance is defined by the number of bits by which the two bit--strings differ 
from each other. Or, in other words, the number of bits which need to be changed to turn 
the language trait of an agent into that of the other one. Figure~\ref{fig:hamming} 
shows that the peak of the histogram of Hamming distances turns from zero (dominance) to one 
(no dominance) at the critical point (which here is located between $m=0.6$ and $m=0.7$).

\section{Discussion and conclusions}
\label{sec:concl}

In our model of language competition a phase transition similar to the one for the competition 
of  different grammars in ref.\cite{Komarova2004} and the one for the competition
of languages of ref.\cite{Stauffer2005} is observed. Nevertheless,
working without fitness improvement for agents better adapted by their languages trait but with a 
sexual population with random mating makes the simulations useful to characterize current 
language competition. The phase of an uniform distribution of the languages (fragmentation) 
suggests a scenario 
where these languages compete equally against each other maintaining a stable state of 
coexistence. In the other state one language dominates clearly the other ones. The dependence 
of the phase transition point on the amount of language interaction as well as 
on the number of competing languages lets us make the following conclusions: There is no optimal 
value of maximum learning age for a language. Its increase shifts the critical point to larger 
values of the mutation rate. Thus a larger maximum 
learning age prevents the evolution of many languages where contact among speakers of different 
languages is strong. 
The model shows also that a larger number of different languages decreases the possibility of 
dominance of one of them and makes the transition jump higher. 
Hence, the extinction of a language in a multilingual system can lead to a transition to the state
where one language will begin to dominate all the other ones.

The phase transition analyzed in the presented model is of first order for simulations 
with many languages as found in the model of ref.\cite{Stauffer2005,Komarova2004}. 
The differences are that in our case, 
the transition is completely independent of the population size.
We do not know if the simplification of a language to a single bit leads to this 
discrepancy or if it is the usage of a model with sexual reproduction. Further analysis has to be
done.

The model brought insight into the mechanisms of language competition but still much has 
to be done. The results are similar to the ones of ref.\cite{Stauffer2005}. Nevertheless,
our model presents another approach to understand language competition. 
The design of an analytical model, for instance by diffusion equations similar to 
the ones used in ref.\cite{Patriarca2004}, 
without fitness would reveal if it is possible to obtain similar results with a mean 
field theory. 
The model presented in this paper can be extended by putting it on a lattice in order to 
investigate how the transition behaves with respect to a geographic distribution of languages. 
Another extension would 
be the substitution of random mating by assortative mating\cite{Kosmidis2005}, 
a concept used frequently in 
the theory of biological speciation, in order to give different priorities to monolingual 
parents and bilingual ones. In order to
understand language invasion, the model can be extended to include social structures.

\section*{Acknowledgements}
I am funded by the DAAD (Deutscher Akademischer Austauschdienst) and 
thank D. Stauffer and S. Moss de Oliveira 
for comments on my manuscript.

\end{document}